\documentclass[a4paper,11pt]{article}
\usepackage{subfigure}
\usepackage{braket}
\usepackage{amsmath}
\usepackage{amssymb}
\usepackage{bm}
\usepackage{pos}
\def\cT{{\cal T}}
\def\rrangle{\rangle\!\rangle}
\def\llangle{\langle\!\langle}
\newcommand*\diff{\mathop{}\!\mathrm{d}}

\title{Quantum simulation of jet evolution in a medium
}

\author*[a]{Wenyang Qian}

\affiliation[a]{Instituto Galego de Fisica de Altas Enerxias (IGFAE), \\
Universidade de Santiago de Compostela, E-15782 Galicia, Spain}

\emailAdd{qian.wenyang@usc.es}

\abstract{
Jets provide one of the primary probes of the quark-gluon plasma produced in ultrarelativistic
heavy ion collisions and the cold nuclear matter explored in deep inelastic scattering experiments. However, despite important developments in the last years, a description of the real-time evolution
of QCD jets inside a medium is still far from complete. 
In our previous work, we have explored
quantum technologies as a promising alternative theoretical laboratory to simulate jet evolution in
QCD matter, to overcome inherent technical difficulties in present calculations. Here, we extend
our previous investigation from the single particle to the multiple particle Fock spaces, taking into
account gluon production. 
Based on the light-front Hamiltonian formalism, we construct a digital quantum circuit that tracks the evolution of a multi-particle jet probe in the presence of a stochastic color background. Using the quantum simulation algorithm, we show the medium-induced modification to the jet evolution in both the momentum broadening and gluon production.

}

\FullConference{HardProbes2023\\
 26-31 March 2023 \\
 Aschaffenburg, Germany\\}

\begin{document}
\maketitle

\section{Introduction}
Jets can resolve the underlying medium at different energy scales and thus offer an optimal probe to study the structure of QCD matter. Their evolution in these environments is characterized by sizeable modifications to the jets' structure~\cite{Casalderrey-Solana:2007knd}. To the present day, the theoretical study of these effects has been mainly constrained to lower orders in perturbation theory, with a limited number of higher-order calculations, where the difficulty is mainly tied to the highly complex multi-particle interference determining parton fragmentation in matter. Recently, novel advances in quantum computing showed us an alternative path to understanding in-medium jet physics. In our preceding work, we provided a quantum simulation protocol to simulate the real-time evolution of a single hard parton in the presence of a stochastic background. In this work, we extend the strategy to include gluon radiation, allowing multiple particles present in the jet probe.

\section{Theoretical methodology}\label{sec:method}

\begin{figure}[!t]
\centering
    \includegraphics[width=0.38\textwidth]{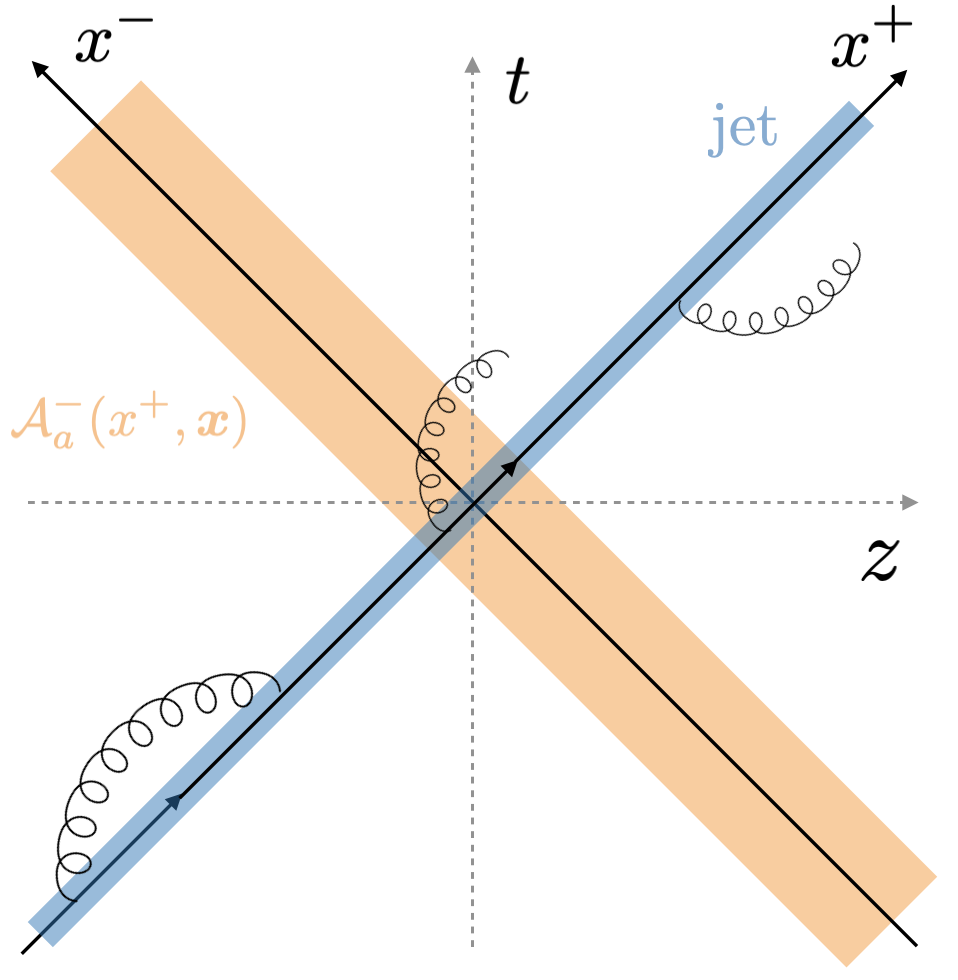}
    \caption{Light-cone diagram for the physical setup.
    }
    \label{fig:lc_diagram}
\end{figure}

We consider the propagation of a highly energetic massless jet with light-front momentum $p=(p^+, p^-, {\bm p})$, moving close to the light cone along the $x^+$ direction.\footnote{The light-front coordinates are defined as \( (x^+, x^-, {\bm x}) \), where \(x^+=x^0+ x^3\) is the light-front time,  \(x^-=x^0-x^3\) the longitudinal coordinate, and  \({\bm x}=(x^1, x^2)\) the transverse coordinates.
} This hard probe evolves in the presence of a dense medium, which is boosted along the $x^-$ direction. The jet-medium interaction occurs over a finite light-front time $x^+$; see Fig.~\ref{fig:lc_diagram} for an illustration. The QCD Lagrangian sets the dynamics of this system in the presence of an external field,
\begin{align}\label{eq:Lagrangian}
 \mathcal{L}=-\frac{1}{4}{F^{\mu\nu}}_a F^a_{\mu\nu}+\overline{\Psi}(i\gamma^\mu  D_\mu -  m_q)\Psi\;,
\end{align}
where $F^{\mu\nu}_a\equiv\partial^\mu C^\nu_a-\partial^\nu C^\mu_a-g f^{abc}C^\mu_b C^\nu_c$ is the field strength tensor, $D^\mu\equiv \partial_\mu +ig C^\mu$ the covariant derivative, where $ C^\mu= A^\mu + \mathcal{A}^\mu$ sums the quantum gauge field $ A^\mu$ and background field $\mathcal{A}^\mu$. The Fock space of the jet is truncated to the leading  two sectors, $\ket{q}$ and $\ket{qg}$, such that
\begin{align}\label{eq:Fock}
    \ket{\psi} = \psi_{q}\ket{q} + \psi_{qg} \ket{qg}\;,
\end{align}
where $\psi_{q}$ and $\psi_{qg}$ represent their respective Fock amplitudes. 
By Legendre transformation, the light-front (LF) Hamiltonian, in light-cone gauge $A^+=\mathcal{A}^+=0$, becomes
\begin{align}\label{eq:Hamiltonian_full}
P^-(x^+)=P_{KE}^- + V_\mathrm{qg}+ V_\mathcal{A}(x^+)\;.
\end{align}
Here, $P_{KE}^- = P_{KE,g}^- + P_{KE,q}^-$ stands for the kinetic energy for the quark and the dynamical gluon.
$V_{qg}$ is the interaction between the quark and gluon, responsible for gluon emission and absorption. Lastly, $V_\mathcal{A}(x^+)$ includes the interaction of the quark and the dynamical gluon with the background field $\mathcal{A}^\mu$. See Ref.~\cite{Li:2021zaw} for more details.

As in our preceding work~\cite{Barata:2022wim}, we take McLerran-Venugopalan (MV) model to describe the background field $\mathcal{A}$, and make use of high energy (eikonal) limit, where $p^+\gg p^\perp,p^-$. Nonetheless, the Hamiltonian method allows us to surpass the formal eikonal limit of $p^+=\infty$. 
The medium color charge density obeys a Gaussian and local correlation function
\begin{align}\label{eq:MV_color_charge}
 \llangle \rho_a(x^+,\bm x)\rho_b(y^+,\bm y) \rrangle =g^2 \mu^2\delta_{ab}\,\delta^{(2)}(\bm x-\bm y)\,\delta(x^+-y^+),
\end{align}
where $\llangle \cdots \rrangle$ denotes the medium configuration average, and $\mu$ the medium strength. The saturation scale $Q_s^2\equiv C_F g^4 \mu^2 L_{\eta}/(2\pi)$ with $C_F= (N_c^2-1)/{(2N_c)}$. The field is solved from the reduced classical Yang-Mills equation with a gluon mass $m_g$ introduced to regularize the infrared divergence.

The time evolution of the quark jet, as a quantum state $\ket{\psi(x^+)}$, follows the time-dependent Schr\"{o}dinger equation, with light-front time ordering operator $\cT_{+}$ in path-ordered form,
\begin{align}~\label{eq:evolution_eq}
 \ket{\psi(x^+)}= U(x^+;0)\ket{\psi(0)}
 \equiv \cT_+ \exp \big[-\frac{i}{2}\int_{0}^{x^+} \diff z^+\, P^-(z^+) \big] \ket{\psi(0)}.
\end{align}
\section{Quantum simulation algorithm}
The digital quantum simulation algorithm involves five generic steps: input, encoding, initial state preparation, time evolution, and measurement.
Here, we extend the algorithm developed in Ref.~\cite{Barata:2022wim} to the $\ket{q}+\ket{qg}$ sectors and highlight on basis encoding and time evolution.

\subsection{Basis encoding}\label{sec:basis_encoding}
We choose the eigenstates of the Hamiltonian $P^-_{KE}$ as the basis states and each single particle state carries five quantum numbers~\cite{Li:2021zaw}, i.e., the LF momenta $\{p^+, p^x, p^y\}$, color $c$, and LF helicity $\lambda$. Here, we describe the encoding for the Fock spaces $\ket{q}+\ket{qg}$ used in this work. 
The complete basis encoding for any basis state $\ket{\beta_{\psi}}$ in the $\ket{q}$ + $\ket{qg}$ Fock sectors is
\begin{align}\label{eq:encoding_complete}
    \ket{\beta_{\psi}} \to \ket{\zeta}\otimes \underbrace{\Big(\ket{p_g^x} \ket{p_g^y}\ket{c_g}\Big)}_{\ket{g}} \otimes \underbrace{\Big(\ket{p_q^x} \ket{p_q^y}\ket{c_q}\Big)}_{\ket{q}}.
\end{align}
In the \textbf{transverse} direction, we use registers $\ket{p_q^x} \ket{p_q^y}$ and $\ket{p_g^x} \ket{p_g^y}$ as two-dimensional periodic lattices to encode for the quark and gluon, Both lattices span a length of $2 L_\perp$ over $2 N_\perp$ sites per dimension. Notably, position and momentum spaces are related by a discrete Fourier Transform ($\mathcal{FT}$), or a quantum $\mathcal{FT}$ on the circuit~\cite{nielsen_chuang_2010}. In the \textbf{color} space where we consider $N_c=2$, we use binary encoding to encode the $N_c=2$ colors for the quark and $N_c^2-1=3$ colors for the gluon. In the \textbf{helicity} space, we make a simplification by considering the non-zero helicity-non-flip term in the massless quark limit, i.e., $\lambda_q=\lambda_g=\uparrow$, and thus no encoding is needed. Lastly, in the \textbf{longitudinal} and \textbf{Fock-occupancy} spaces, we use a combined register $\ket{\zeta}$. Speaking in terms of discretized total momentum quanta $K$, we use $\zeta=0$ to encode the $\ket{q}$ state with $k_Q^+ = K$; and $\zeta=\{1,2,\cdots K-1/2\}$ to encode the $\ket{qg}$ states with $k_g^+=\{1,2,\cdots,K-1/2\}$ and consequently $k_q^+=K-k_g^+$.

In conclusion, the number of qubits for two Fock sectors is $n_Q = 4\log_2{N_\perp}+\log{\lceil K \rceil} + 7$  with a transverse $N_\perp$ and longitudinal $K$. For $n$ particles, it scales with $(2n)\log(N_\perp)$ qubits, a logarithmic reduction compared to classical calculations, though efficient simulation is still required.

\begin{figure}[!t]
\centering
    \includegraphics[width=0.8\textwidth]{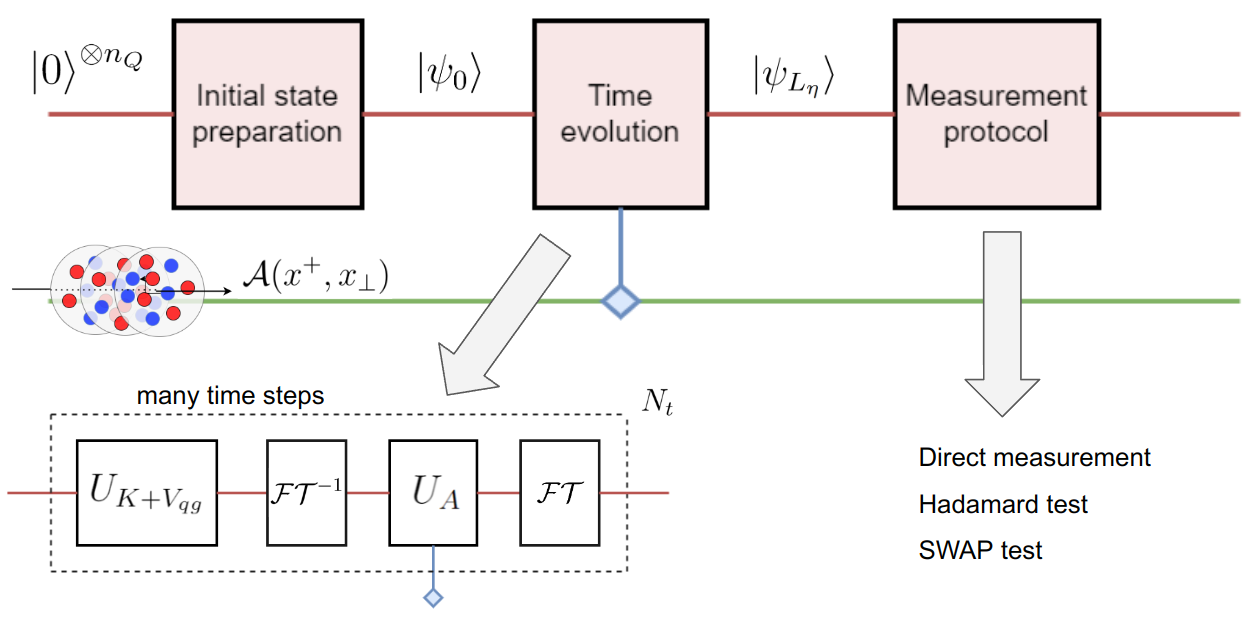}
    \caption{Quantum simulation algorithm for jet in a medium. Notably, $\mathcal{FT}: x\mapsto p$ and $\mathcal{FT}^{-1}: p\mapsto x$.
    }
    \label{fig:qc}
\end{figure}

\subsection{Unitary time evolution}\label{sec:gate_encoding}

Naturally, the evolution of the quantum state is unitary. We use the product formula to split the evolution as in Eq.~\eqref{eq:evolution_eq} along the $x^+$ direction into $N_t$ total time steps, 
\begin{align}~\label{eq:evolution_product}
 U(x^+=L_\eta;0)
=& \prod_{k=1}^{N_t} U(x^+_k;x^+_{k-1})\;,
\end{align}
where $x^+_k=k\, L_\eta/N_t$ is the intermediate time with duration $\delta x^+=L_\eta/N_t$. Notably, the Hamiltonian is time-dependent for stochastic fields $\mathcal{A}^-_a(x^+,\bm x)$. Using the MV model, the medium is sliced into $N_\eta$ layers along $x^+$, and within each time step $\delta x^+$, the evolution operator is
\begin{align}\label{eq:unitary_mixed}
\begin{split}
   \quad  U( x_k^++\delta x^+; x^+_k) 
    \approx  [\mathcal{FT}] \exp \bigg\{-i\delta x^+ \left[{V}_{\mathcal{A}}(x_k^+)\right]\bigg \} [\mathcal{FT}^{-1}]
    \times\exp \bigg\{-i\delta x^+ \left[ {K}+ {V}_{qg} \right]\bigg \} ,
    \end{split}
\end{align}
Here, the evolution of the kinetic energy and gluon emission, i.e., ${K}+{V}_{qg}$, is in momentum space; whereas the evolution of the medium, i.e., ${V}_{\mathcal{A}}$, is in position space. 
Since $\mathcal{A}(x_k^+, \bm x)$ is diagonal in position, this alternating mixed-space evolution (see Fig.~\ref{fig:qc}) is more economical than direct exponentiation, potentially extending to larger lattices and higher Fock sectors.

\section{Results}

\begin{figure}[!t]
\centering
    \includegraphics[width=0.42\textwidth]{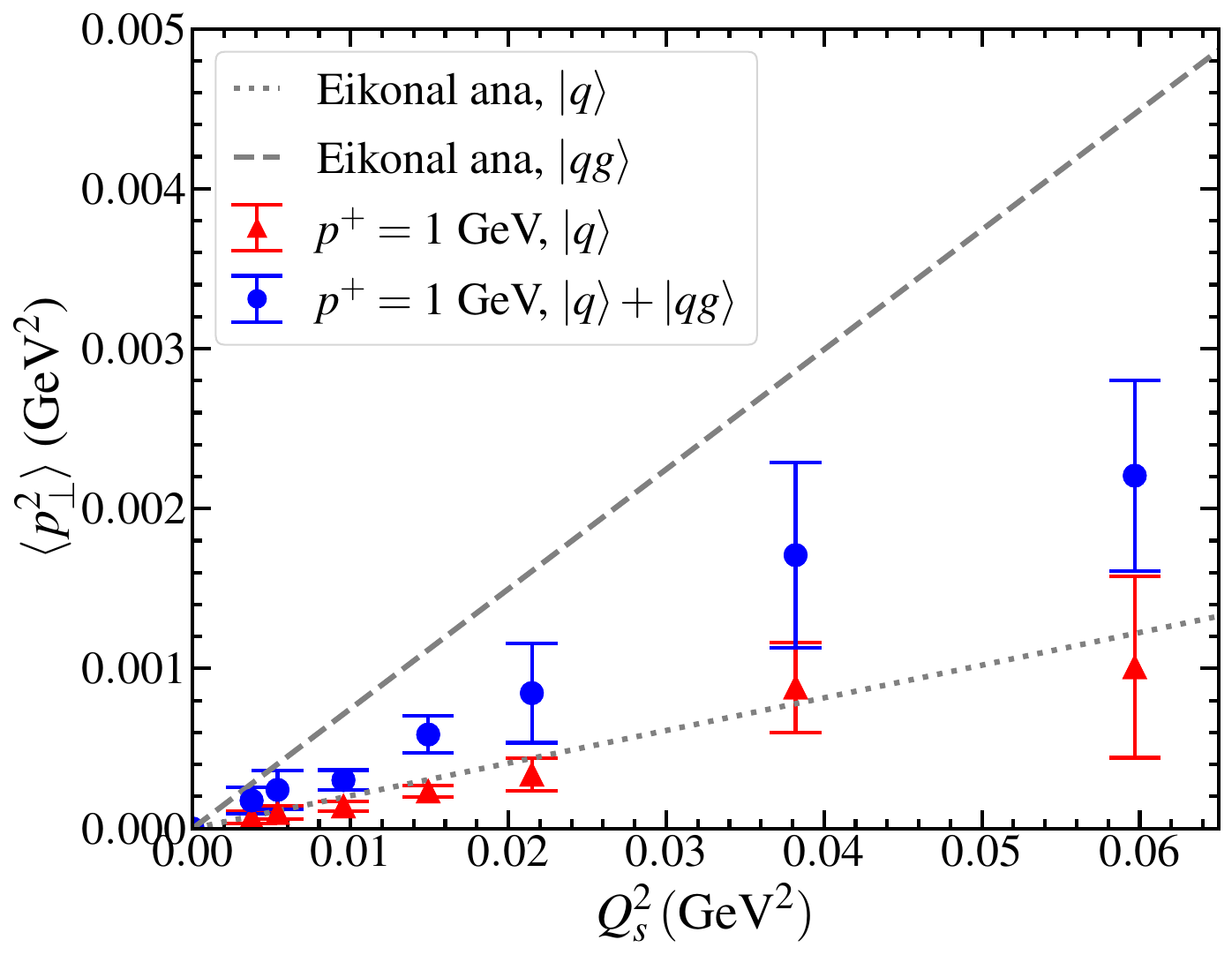}
    \quad
    \includegraphics[width=0.42\textwidth]{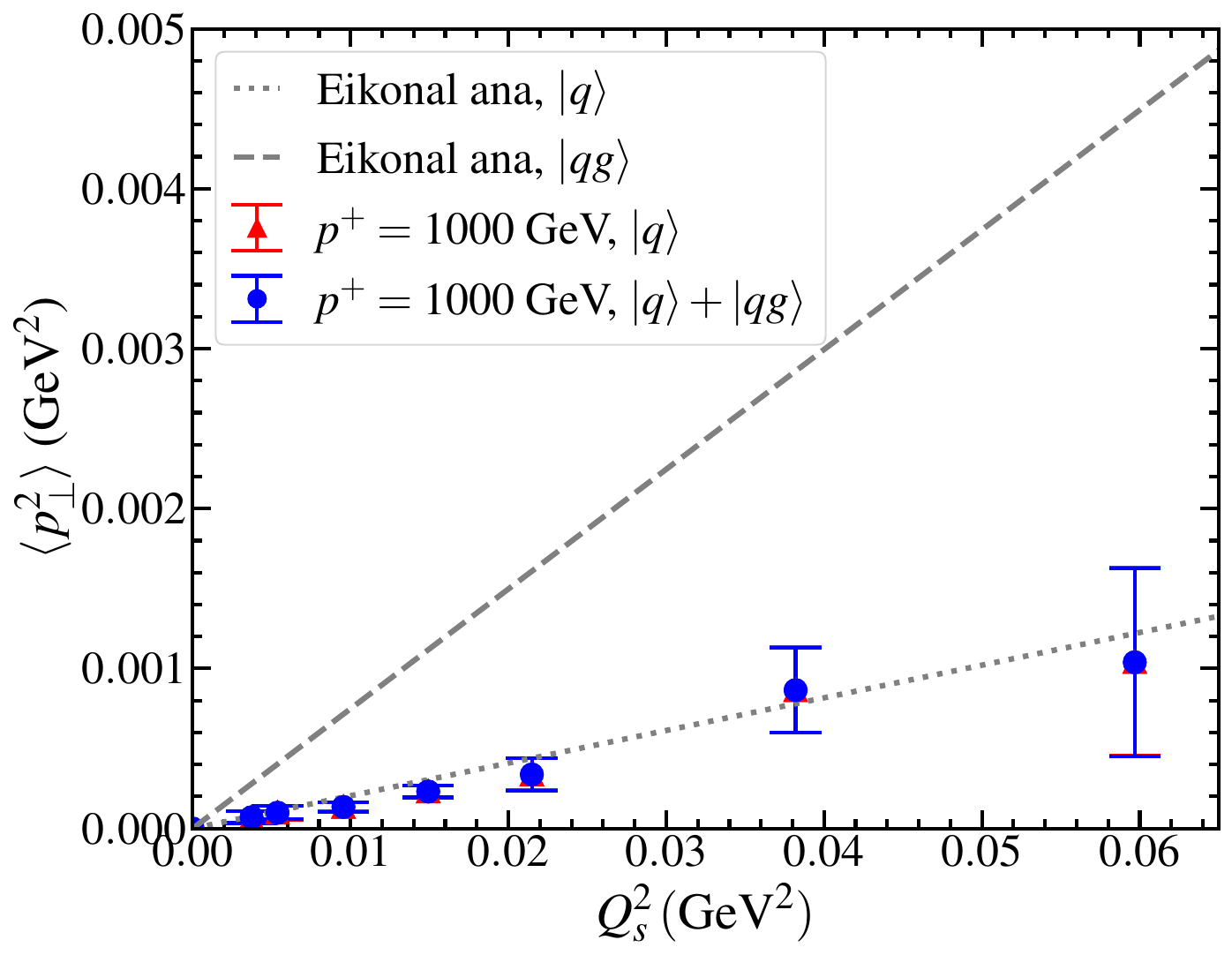}
    \caption{Momentum broadening $\braket{p_\perp^2}$ on the saturation scale $Q_s$.
    }
    \label{fig:p2_full}
\end{figure}
\begin{figure}[!t]
    \centering
    \quad
    \includegraphics[width=0.42\textwidth]{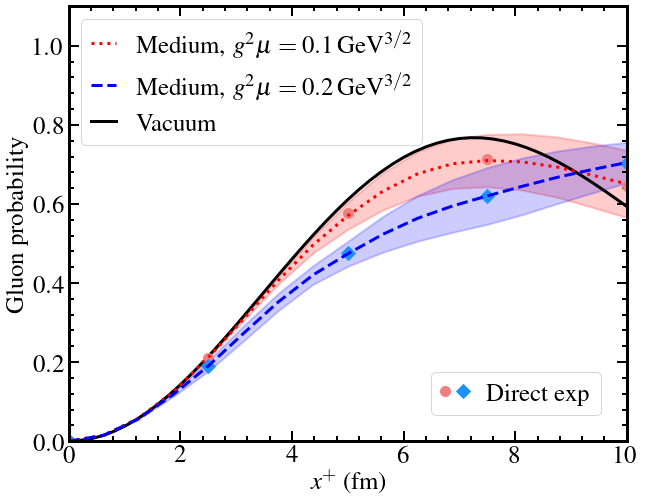}
    \quad
    \includegraphics[width=0.42\textwidth]{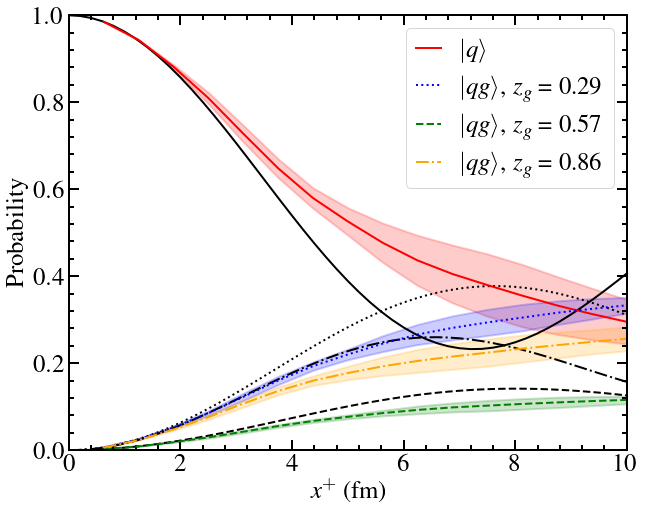}    
    \caption{Gluon production in real-time evolution, with its momentum fraction $z_g\equiv p^+_g/P^+=\zeta/K$. 
    }
    \label{fig:evo_med}
\end{figure}

Using the quantum simulation algorithm, we study the jet momentum broadening and the gluon production for various medium strengths. The quark state is initialized as a zero transverse momentum state with superposition in color. For the simulation, we take the transverse lattice of $N_\perp=1$ and total longitudinal momentum quanta $K=3.5$, with $n_Q = 9$ qubits in total. For the physical setup, we take $L_{\perp}=6.4$~fm and the medium duration $L_{\eta}=10$ fm. We use the ideal shot-based {\tt QasmSimulator} with 819200 shots from \texttt{Qiskit} to measure observables directly, and the uncertainties on our plots are related to the medium field fluctuations.

We reconstruct the total transverse momentum of the jet, as
$\braket{p^2_\perp}
= \braket{\psi(L_\eta)| \hat{p}^2_\perp|\psi(L_\eta)}
$, in Fig.~\ref{fig:p2_full}, via extracting the final jet probability distribution. With a zero initial state, $\braket{p_\perp^2}$ indicates the broadening effect exclusively from the medium. The $\braket{p^2}$ results are presented at various saturation scales $Q_s$ where the eikonal approximation is relaxed with finite energy, $p^+=1, 1000$ GeV. Notably, the $p^+=1000$ GeV result overlaps with the eikonal limit for the $\ket{q}$, for both kinetic and gluon emission parts are highly suppressed.
By contrast, the $p^+= 1~\mathrm{GeV}$ result of $\ket{q}+\ket{qg}$ lies between the two eikonal limits, suggesting non-eikonal effects due to gluon emission, which is, intuitively, owing to the enlarged phase space with the inclusion of $\ket{qg}$ sector.

Furthermore, with this setup, we can also study the real-time evolution of in-medium gluon production, as in Fig.~\ref{fig:evo_med}. Here, we see the total gluon probability $\mathcal{P}_{qg}$ (left panel) and its longitudinal distributions (right panel), both as results from the quantum decoherence among different modes. However, since our momentum lattice space is limited, it is hard to conclude whether the medium induces or suppresses radiation production. Larger lattices are needed to fully justify the connection to the Landau-Pomeranchuk-Migdal effect.

\section{Conclusion and Outlook}
In this work, we implemented a digital quantum simulation protocol to study the real-time evolution of a multi-particle jet in QCD matter, based on light-front Hamiltonian formalism. There are several paths to take from here. Extensions of the algorithm to include higher Fock sectors, such as $\ket{qgg} $, are underway,  allowing us to perform numerical calculations beyond known analytical results. Another interesting avenue to explore is the transition of the final partonic jet into a hadronic state, using a variational pion state on the circuit~\cite{Qian_PRR}. It is also interesting to study the thermalization of the medium and the medium response on the quantum circuit in the future.

\section*{Acknowledgments}
We are grateful to J. Barata, X. Du, M. Li, and C. A. Salgado, who have made important contributions to this work, i.e., Ref.~\cite{Barata:2023clv}.
WQ is supported by Xunta de Galicia (Centro singular de investigacion de Galicia accreditation 2019-2022), European Union ERDF, the “Maria de Maeztu” Units of Excellence program under project CEX2020-001035-M, the Spanish Research State Agency under project PID2020-119632GB-I00, European Research Council under project ERC-2018-ADG-835105 YoctoLHC, and European Union's Marie Skłodowska-Curie Actions Postdoctoral Fellowships 2022 under Grant Agreement No. 101109293.

\bibliographystyle{plain}
\bibliography{my_bib.bib}

\end{document}